\documentclass[11pt, twocolumn,twoside]{IEEEtran}

\usepackage{graphicx} 
\usepackage{amssymb}
\usepackage{verbatim}
\usepackage{graphicx}
\usepackage{amsmath}
\usepackage{url}
\usepackage{amsthm, bm}
\usepackage{enumerate}
\usepackage[square,sort&compress,comma,numbers]{natbib}
\usepackage{cleveref}
\usepackage{float}
\usepackage{dblfloatfix}
\usepackage[font=small,labelfont=bf]{caption}

\DeclareMathOperator*{\argmin}{arg\,min}
\newcommand{\dd}[1]{\,{\rm d} #1}  

\makeatletter
\def\th@plain{%
  \thm@notefont{}
  \itshape 
}
\def\th@definition{%
  \thm@notefont{}
  \normalfont 
}
\makeatother

\begin{document}
\title{Deep Learning Methods for Parallel Magnetic Resonance Image Reconstruction}

\author{Florian~Knoll, Kerstin~Hammernik, Chi~Zhang,~\IEEEmembership{Student Member,~IEEE,} Steen~Moeller, Thomas~Pock, Daniel~K.~Sodickson,~and~Mehmet~Ak\c{c}akaya,~\IEEEmembership{Member,~IEEE}
	\vspace{-0.3cm}
	\thanks{F. Knoll and D. K. Sodickson are with the Center for Biomedical Imaging, Department of Radiology, New York University. K. Hammernik and T. Pock are with the Institute of Computer Vision and Graphics, Graz University of Technology. C. Zhang and  M. Ak\c{c}akaya are with the Department of Electrical and Computer Engineering, and Center for Magnetic Resonance Research, University of Minnesota, Minneapolis, MN. S. Moeller is with the Center for Magnetic Resonance Research, University of Minnesota, Minneapolis, MN. 
	e-mails: florian.knoll@nyumc.org, hammernik@icg.tugraz.at, zhan4906@umn.edu, moeller@cmrr.umn.edu, pock@icg.tugraz.at, daniel.sodickson@nyumc.org, akcakaya@umn.edu. 
	This work was partially supported by NIH R01EB024532, NIH R00HL111410, NIH P41EB015894, NIH P41EB027061, NIH P41EB017183, NSF CAREER CCF-1651825}}
	
\maketitle

\begin{abstract}
 Following the success of deep learning in a wide range of applications, neural network-based machine learning techniques have received interest as a means of accelerating magnetic resonance imaging (MRI). A number of ideas inspired by deep learning techniques from computer vision and image processing have been successfully applied to non-linear image reconstruction in the spirit of compressed sensing for both low dose computed tomography and accelerated MRI. The additional integration of multi-coil information to recover missing k-space lines in the MRI reconstruction process, is still studied less frequently, even though it is the de-facto standard for currently used accelerated MR acquisitions. This manuscript provides an overview of the recent machine learning approaches that have been proposed specifically for improving parallel imaging. A general background introduction to parallel MRI is given that is structured around the classical view of image space and k-space based methods. Both linear and non-linear methods are covered, followed by a discussion of recent efforts to further improve parallel imaging using machine learning, and specifically using artificial neural networks. Image-domain based techniques that introduce improved regularizers are covered as well as k-space based methods, where the focus is on better interpolation strategies using neural networks. Issues and open problems are discussed as well as recent efforts for producing open datasets and benchmarks for the community.
\end{abstract}

\IEEEpeerreviewmaketitle	

\section{Introduction}
During recent years, there has been a substantial increase of research activity in the field of medical image reconstruction. One particular application area is the acceleration of Magnetic Resonance Imaging (MRI) scans. This is an area of high impact, because MRI is the leading diagnostic modality for a wide range of exams, but the physics of its data acquisition process make it inherently slower than modalities like X-Ray, Computed Tomography or Ultrasound. Therefore, the shortening of scan times has been a major driving factor for routine clinical application of MRI.

One of the most important and successful technical developments to decrease MRI scan time in the last 20 years was parallel imaging ~\cite{Sodickson1997,Pruessmann1999,Griswold2005}. Currently, essentially all clinical MRI scanners from all vendors are equipped with parallel imaging technology, and it is the default option for a large number of scan protocols. As a consequence, there is a substantial benefit of using multi-coil data for machine learning based image reconstruction. Not only does it provide a complementary source of acceleration that is unavailable when operating on single channel data, or on the level of image enhancement and post-processing, it also is the scenario that ultimately defines the use-case for accelerated clinical MRI, which makes it a requirement for clinical translation of new reconstruction approaches. The drawback is that working with multi-coil data adds a layer of complexity that creates a gap between cutting edge developments in deep learning~\cite{LeCun2015} and computer vision, where the default data type are images. The goal of this manuscript is to bridge this gap by providing both a comprehensive review of the properties of parallel MRI, together with an introduction how current machine learning methods can be used for this particular application.
		
\begin{figure*}[!t]
	\centering
	\includegraphics[width = 6.5in]{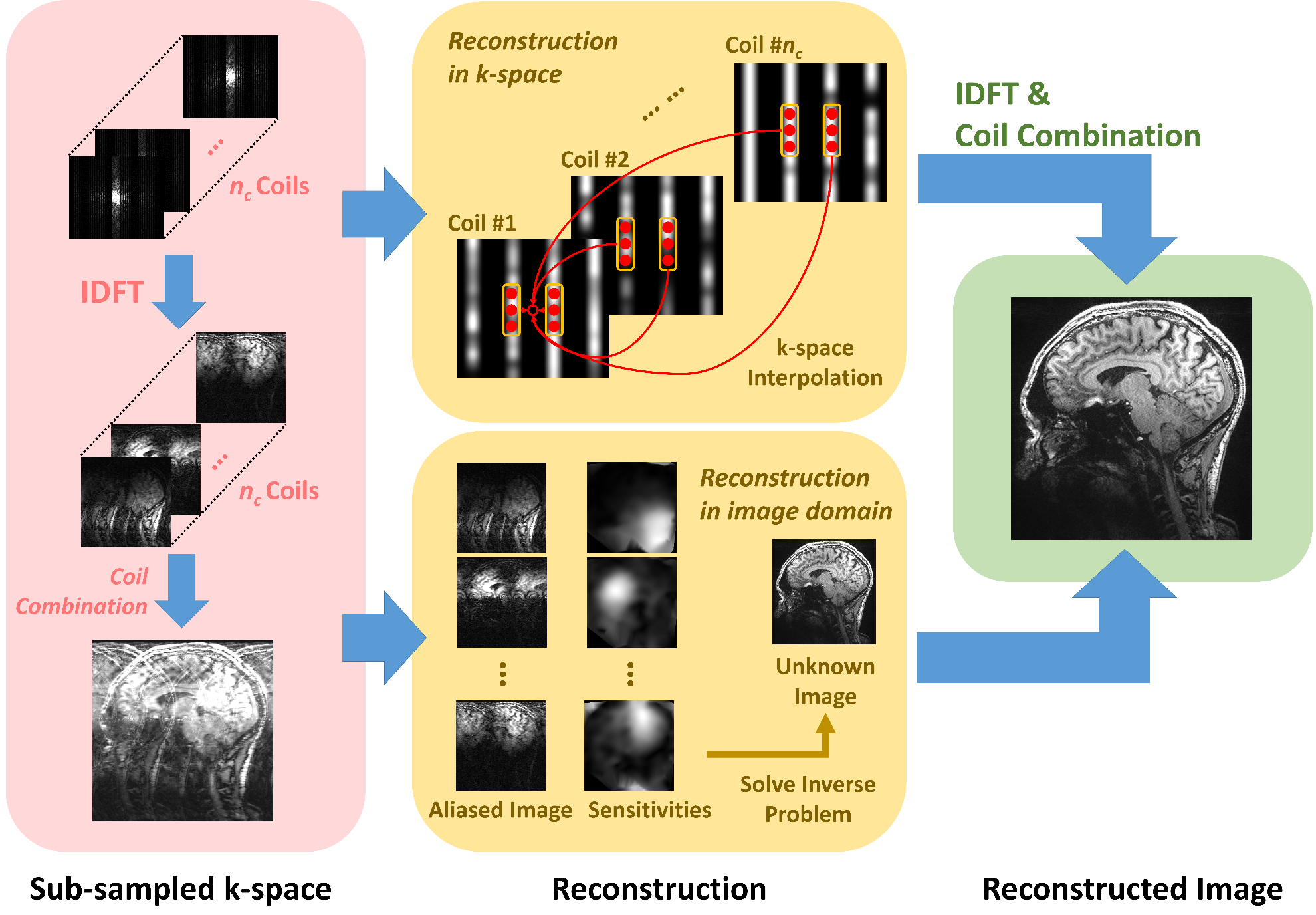}
 	\caption{In k-space based parallel imaging methods, missing data is recovered first in k-space, followed by an inverse Fourier transform and combination of the individual coil elements. In image space based parallel imaging, the Fourier transform is performed as the first step, followed by coil sensitivity based removal of the aliasing artifacts from the reconstructed image by solving an inverse problem.}
 	\vspace{-.3cm}
 	\label{fig:recon_PI_image_kspace}
\end{figure*}		
		
\subsection{Background on multi-coil acquisitions in MRI}
The original motivation behind phased array receive coils~\cite{Roemer1990} was to increase the SNR of MR measurements. These arrays consist of $n_c$ multiple small coil elements, where an individual coil element covers only a part of the imaging field of view. These individual signals are then combined to form a single image of the complete field of view. The central idea of all parallel imaging methods is to complement spatial signal encoding of gradient fields with information about the spatial position of these multiple coil elements. For multiple receiver coils, the MR signal equation can be written as follows
\begin{equation}
\label{eq:signal_multiplecoils}
f_j(k_x,k_y) = \int \limits_{-\infty}^{\infty}	\int \limits_{-\infty}^{\infty} u(x,y) c_j(x,y) e^{-i (k_x x + k_y y)}\dd{x} \dd{y}.
\end{equation}
In \Cref{eq:signal_multiplecoils}, $f_j$ is the MR signal of coil $j=1,\ldots,n_c$, $u$ is the target image  to be reconstructed, and $c_j$ is the corresponding coil sensitivity. Parallel imaging methods use the redundancies in these multi-coil acquisitions to reconstruct undersampled k-space data. After discretization, this undersampling is described in matrix-vector notation by
 \begin{equation}
  \mathbf{f}_j = \mathbf{F}_{\Omega}\mathbf{C}_j\mathbf{u} + \mathbf{n}_j,
 \end{equation}
 where $\mathbf{u}$ is the image to be reconstructed, $\mathbf{f}_j$ is the acquired k-space data in the $j^\textrm{th}$ coil, $\mathbf{C}_j$ is a diagonal matrix containing 
 the sensitivity profile of the receiver coil \cite{Pruessmann1999}, $\mathbf{F}_{\Omega}$ is a partial Fourier sampling operator that samples locations ${\Omega},$ and $\mathbf{n}_j$ is measurement noise in the $j^{\textrm{th}}$ coil.

Historically, parallel imaging methods were put in two categories: Approaches that operate in image domain, inspired by the sensitivity encoding (SENSE) method~\cite{Pruessmann1999} and approaches that operated in k-space, inspired by simultaneous acquisition of spatial harmonics (SMASH)~\cite{Sodickson1997} and generalized autocalibrating partial parallel acquisition (GRAPPA)~\cite{Griswold2002}. This is conceptually illustrated in \Cref{fig:recon_PI_image_kspace}. While these two schools of thought are closely related~\cite{Kholmovski,Uecker2014}, we organized this document according to these classic criteria for historical reasons.




\section{Classical parallel imaging in image space}
\label{sec:imagespace_PI}
Classical parallel imaging in image space follows the SENSE method~\cite{Pruessmann1999}, which can be identified by two key features. First, the elimination of the aliasing artifacts is performed in image space after the application of an inverse Fourier transform. Second, information about receive coil sensitivities is obtained via precomputed, explicit coil sensitivity maps from either a separate reference scan or from a fully sampled block of data at the center of k-space (all didactic experiments that are shown in this manuscript follow the latter approach). More recent approaches jointly estimate coil sensitivity profiles during the image reconstruction process~\cite{Ying2007,Uecker2008}, but for the rest of this manuscript, we assume that sensitivity maps were precomputed. The reconstruction in image domain in \Cref{fig:recon_PI_image_kspace} shows three example undersampled coil images, corresponding coil sensitivity maps and the final reconstructed images from a brain MRI dataset. The coil sensitivities were estimated using ESPIRiT~\cite{Uecker2014}.


MRI reconstruction in general and parallel imaging in particular can be formulated as an inverse problem. This provides a general framework that allows easy integration of the concepts of regularized and constrained image reconstruction as well as machine learning that are discussed in more detail in later sections. \Cref{eq:signal_multiplecoils} can be discretized and then written in matrix-vector form:
\begin{equation}
	\label{eq:forward_problem}
	{\mathbf{f} = \mathbf{E}\mathbf{u} +\mathbf{n}},
\end{equation}
where $\mathbf{f}$ contains all k-space measurement data points and $\mathbf{E}$ is the forward encoding operator that includes information about the sampling trajectory and the receive coil sensitivities and $\mathbf{n}$ is measurement noise. The task of image reconstruction is to recover the image $\mathbf{u}$. In classic parallel imaging one generally operates under the condition that the number of receive elements is larger than the acceleration factor. Therefore, \Cref{eq:forward_problem} corresponds to an over-determined system of equations. However, the rows of $\mathbf{E}$ are linearly dependent because individual coil elements do not measure completely independent information. Therefore the inversion of $\mathbf{E}$ is an ill-posed problem, which can lead to severe noise amplification, described via the g-factor in the original SENSE paper~\cite{Pruessmann1999}. \Cref{eq:forward_problem} is usually solved in an iterative manner, which is the topic of the following sections.
 
\subsection{Overview of conjugate gradient SENSE (CG-SENSE)}
The original SENSE approach is based on equidistant or uniform Cartesian k-space sampling, where the aliasing pattern is defined by a point spread function that has a small number of sharp equidistant peaks. This property leads to a small number of pixels that are folded on top of each other, which allows a very efficient implementation~\cite{Pruessmann1999}. When using alternative k-space sampling strategies like non-Cartesian acquisitions or random undersampling, this is no longer possible and image reconstruction requires a full inversion of the encoding matrix in \Cref{eq:forward_problem}. This operation is demanding both in terms of compute and memory requirements (the dimensions of $\mathbf{E}$ are the total number of acquired k-space points times $N^2$ where $N$ is the size of the image matrix that is to be reconstructed), which lead to the development of iterative methods, in particular the CG-SENSE method introduced by Pruessmann et al. as a follow up of the original SENSE paper~\cite{Pruessmann2001}. In iterative image reconstruction the goal is to find a $\hat{\mathbf{u}}$ that is a minimizer of the following cost function, which corresponds to the quadratic form of the system in \Cref{eq:forward_problem}:
\begin{equation}
  \label{eq:inverse_leastSquares}
  \hat{\mathbf{u}} \in \argmin\limits_{\mathbf{u}} \frac{1}{2} \| \mathbf{E} \mathbf{u} - \mathbf{f} \|_2^2.
\end{equation}
In standard parallel imaging, $\mathbf{E}$ is linear and \Cref{eq:inverse_leastSquares} is a convex optimization problem that can be solved with a large number of numerical algorithms like gradient descent, Landweber iterations~\cite{Landweber1951}, primal-dual methods~\cite{Pock_PD_2010} or the alternating direction method of multipliers (ADMM) algorithm~\cite{Boyd2011} (a detailed review of numerical methods is outside the scope of this article). 
In the original version of CG-SENSE~\cite{Pruessmann2001}, the conjugate gradient method~\cite{Hestenes1952} is employed.  However, since MR k-space data are corrupted by noise, it is common practice stop iterating before theoretical convergence is reached, which can be seen as a form of regularization.
Regularization can be also incorportated via additional constraints in \Cref{eq:inverse_leastSquares}, which will be covered in the next section.

As a didactic example for this manuscript, we will use a single slice of a 2D coronal knee exam to illustrate various reconstruction approaches. This data were acquired on a clinical 3T system (Siemens Skyra) using a 15 channel phased array knee coil. A turbo spin echo sequence was used with the following sequence parameters: TR=2750ms, TE=27m, echo train length=4, field of view 160mm$^2$ in-plane resolution 0.5mm$^2$, slice thickness 3mm.
Readout oversampling with a factor of 2 was used, and all images were cropped in the frequency encoding direction (superior-inferior) for display purposes. In the spirit of reproducible research, data, sampling masks and coil sensitivity estimations that were used for the numerical results in this manuscript are available online\footnote{\url{https://app.globus.org/}: Endpoint: NYULH Radiology Reconstruction Data, coronal pd data, subject 17, slice 25.}. 
\Cref{fig:knee_reconstructions} shows an example of a retrospectively undersampled CG-SENSE reconstruction with an acceleration factor of 4 for the data from \Cref{fig:knee_reconstructions}. Early stopping was employed by setting the numeric tolerance of the iteration to $5\cdot10^{-5}$, which resulted in the the algorithm stopping after 14 CG iterations.

\subsection{Nonlinear regularization and compressed sensing} \label{sec:2c}

\Cref{eq:inverse_leastSquares} can be extended by including a-priori knowledge via additional penalty terms, which results in a constrained optimization problem defined in \Cref{eq:inverse_regularization}, which forms the cornerstone of almost all modern MRI reconstruction methods
\begin{equation}
  \label{eq:inverse_regularization}
  \hat{\mathbf{u}} \in \argmin\limits_{\mathbf{u}} \frac{1}{2}  \| \mathbf{E} \mathbf{u} - \mathbf{f} \|_2^2 + \sum_{i} \lambda_i \Psi_i(\mathbf{u}).
\end{equation}
Here, $\Psi_i$ are dedicated regularization terms and $\lambda_i$ are regularization parameters that balance the trade-off between data fidelity and prior. Since the introduction of compressed sensing~\cite{Candes2006, Donoho2006} and its adoption for MRI~\cite{Lustig2007,Block2007,Lustig2008a}, nonlinear regularization terms, in particular $\ell_1$-norm based ones, are popular in image reconstruction and are commonly used in parallel imaging~\cite{Block2007,Lustig2010,Knoll2011,Knoll2012, Akcakaya2011, Akcakaya2014, Jung2009}. The goal of these regularization terms is to provide a separation between the target image that is to be reconstructed from the aliasing artifacts that are introduced due to an undersampled acquisition. Therefore, they are usually designed in conjunction with a particular data sampling strategy. The classic formulation of compressed sensing in MRI~\cite{Lustig2007} is based on sparsity of the image in a transform domain (Wavelets are a popular choice for static images) in combination with pseudo-random sampling, which introduces aliasing artifacts that are incoherent in the respective domain. For dynamic acquisitions where periodic motion is encountered, sparsity in the Fourier domain common choice~\cite{Gamper2008}. Total Variation based methods have been used successfully in combination with radial~\cite{Block2007} and spiral~\cite{Valvano2016} acquisitions as well as in dynamic imaging~\cite{Feng2013}. More advanced regularizers based on low-rank properties have also been utilized \cite{Lingala2011}.

\Cref{fig:knee_reconstructions} shows an example of a nonlinear combined parallel imaging and compressed sensing reconstruction with a Total Generalized Variation~\cite{Knoll2011} constraint. The regularization parameter $\lambda$ was set to $2.5\cdot10^{-5}$ 
and the reconstruction was using 1000 primal-dual~\cite{Pock_PD_2010} iterations. The used equidistant sampling was chosen for consistency with the other reconstruction methods is not optimal for the incoherence condition in compressed sensing. Nevertheless, the nonlinear regularization still provides a superior reduction of aliasing artifacts and noise suppression in comparison to the CG-SENSE reconstruction from the last section. 
						
\section{Classical parallel imaging in k-space}

Parallel imaging reconstruction can also be formulated in k-space as an interpolation procedure. The initial connections between the SENSE-type image-domain inverse problem approach and k-space interpolation has been made more than a decade ago \cite{Kholmovski}, where it was noted that the forward model in \Cref{eq:forward_problem} can be restated in terms of the Fourier transform, ${\bm \kappa}$ of the combined image, $\mathbf{u}$ as 
\begin{equation}
    \mathbf{f} = \mathbf{A}\mathbf{F}^*{\bm \kappa} \triangleq \mathbf{G}_{\text{acq}} {\bm \kappa},
\end{equation}
where $\mathbf{f}$ corresponds to the acquired k-space lines across all coils, and $\mathbf{G}_{\text{acq}}$ is a linear operator. Similarly, the unacquired k-space lines across all coils can be formulated using 
\begin{equation}
    \mathbf{f}_{\text{unacq}} = \mathbf{G}_{\text{unacq}} {\bm \kappa}
\end{equation}
Combining these two equations yield
\begin{equation}
    \mathbf{f}_{\text{unacq}} = \mathbf{G}_{\text{unacq}} \mathbf{G}_{\text{acq}}^{\dagger} \mathbf{f}.
\end{equation}
Thus, the unaccquired k-space lines across all coils can be interpolated based on the acquired lines across all coils, assuming the pseudo-inverse, $\mathbf{G}_{\text{acq}}^{\dagger}$, of $\mathbf{G}_{\text{acq}}$ exists \cite{Kholmovski}. Thus, the main difference between the k-space parallel imaging methods and the aforementioned image domain parallel imaging techniques is that the former produces k-space data across all coils at the output, whereas the latter typically produces one image that combines the information from all coils.

\subsection{Linear k-space interpolation in GRAPPA}
The most clinically used k-space reconstruction method for parallel imaging is GRAPPA, which uses linear shift-invariant convolutional kernels to interpolate missing k-space lines using uniformly-spaced acquired k-space lines \cite{Griswold2002}. For the $j^\textrm{th}$ coil k-space data, ${\kappa}_j$, we have
\begin{align}
    {\kappa}_j&(k_x, k_y - m\Delta k_y)  \nonumber \\
    &= \sum_{c=1}^{n_c} \sum_{b_x = -B_x}^{B_x} \sum_{b_y = -B_y}^{B_y} g_{j,m} (b_x,b_y,c) \nonumber \\
    &\quad\quad \quad \quad \kappa_c(k_x - b_x \Delta k_x,k_y - Rb_y \Delta k_y),
\end{align}
where $R$ is the acceleration rate of the uniformly undersamped acquisition; $m \in \{1, \dots, R-1\}$; $g_{j,m}(b_x,b_y,c)$ are the linear convolutional kernels for estimating the $m^\textrm{th}$ spacing location in $j^\textrm{th}$ coil; $n_c$ is the number of coils; and $B_x$, $B_y$ are parameters determined from the convolutional kernel size. A high-level overview of such interpolation is shown in the reconstruction in k-space section of \Cref{fig:recon_PI_image_kspace}.

Similar to the coil sensitivity estimation in SENSE-type reconstruction, the convolutional kernels, $g_{j,m}(b_x,b_y,c)$ are estimated for each subject, from either a separate reference scan or from a fully-sampled block of data at the center of k-space, called autocalibrating signal (ACS) \cite{Griswold2002}. A sliding window approach is used in this calibration region to identify the fully-sampled acquisition locations specified by the kernel size and the corresponding missing entries. The former, taken across all coils, is used as rows of a calibration matrix, $\mathbf{A}$; while the latter, for a specific coil, yields a single entry in the target vector, $\mathbf{b}$. Thus for each coil $j$ and missing location $m \in \{1, \dots, R-1\}$, a set of linear equations are formed, from which the vectorized kernel weights $g_{j,m}(b_x,b_y,c)$, denoted $\mathbf{g}_{j,m}$, are estimated via least squares, as $\mathbf{g}_{j,m}\in \arg \min_{\mathbf{g}} ||{\mathbf{b}- \mathbf{Ag}}||_2^2$. GRAPPA has been shown to have several favorable properties compared to SENSE, including lower g-factors, sometimes even less than unity at certain parts of the image \cite{Robson2008}, and more smoothly varying g-factor maps \cite{Breuer2009}. Furthermore, k-space interpolation is often less sensitive to motion \cite{Breuer2005}. Due to these favorable properties, GRAPPA has found utility in multiple large-scale projects, such as the Human Connectome Project \cite{Ugurbil2013}. 


\subsection{Advances in k-space interpolation methods}

Though GRAPPA is widely used in clinical practice, it is a linear method that suffers from noise amplification based on the coil geometry and the acceleration rate \cite{Pruessmann1999}. Therefore, several alternative strategies have been proposed in the literature to reduce the noise in reconstruction.

Iterative self-consistent parallel imaging reconstruction (SPIRiT) is a strategy for enforcing self-consistency among the k-space data in multiple receiver coils by exploiting correlations between neighboring k-space points \cite{Lustig2010}. Similar to GRAPPA, SPIRiT also estimates a linear shift-invariant convolutional kernel from ACS data. In GRAPPA, this convolutional kernel used information from acquired lines in a neighborhood to estimate a missing k-space point. In SPIRiT, the kernel includes contributions from all points, both acquired and missing, across all coils for a neighborhood around a given k-space point. 
The self-consistency idea suggests that the full k-space data should remain unchanged under this convolution operation. SPIRiT objective function also includes a term that enforces consistency with the acquired data, where the undersampling can be performed with arbitrary patterns, including random patterns that are typically employed in compressed sensing \cite{Block2007, Lustig2007}.
Additionally, this formulation allows incorporation of regularizers, for instance based on transform-domain sparsity, in the objective function 
to reduce reconstruction noise via non-linear processing \cite{Lustig2010}
. Furthermore, SPIRiT has facilitated the connection between coil sensitivities used in image-domain parallel imaging methods and the convolutional kernels used in k-space methods via a subspace analysis \cite{Uecker2014}.

An alternative line of work utilizes non-linear k-space interpolation for estimating missing k-space points for uniformly undersampled parallel imaging acquisitions \cite{Chang2012}. It was noted that during GRAPPA calibration, both the regressand and the regressor have errors in them due to measurement noise in the acquisition of calibration data, which leads to a non-linear relationship in the estimation. Thus, the reconstruction method, called non-linear GRAPPA, uses a kernel approach to map the data to a higher-dimensional feature space, where linear interpolation is performed, which also corresponds to a non-linear interpolation in the original data space. The interpolation function is estimated from the ACS data, although this approach typically required more ACS data than GRAPPA \cite{Chang2012}. This method was shown to reduce reconstruction noise compared to GRAPPA. Note that non-linear GRAPPA, through its use of the kernel approach, is a type of machine learning approach, though the non-linear kernel functions were empirically fixed a-priori and not learned from data.

\subsection{k-space reconstruction via low-rank matrix completion}
While k-space interpolation methods remain the prevalent method for k-space parallel imaging reconstruction, there has been recent efforts on recasting this type of reconstruction as a matrix completion problem. Simultaneous autocalibrating and k-space estimation (SAKE) is an early work in this direction, where local neighborhoods in k-space across all coils are restructured into a matrix with block Hankel form \cite{Shin2014}. Then low-rank matrix completion is performed on this matrix, subject to consistency with acquired data, enabling k-space parallel imaging reconstruction without additional calibration data acquisition. Low-rank matrix modeling of local k-space neighborhoods (LORAKS) is another method exploiting similar ideas, where the motivation is based on utilizing finite image support and image phase constraints instead of correlations across multiple coils \cite{Haldar2014}. This method was later extended to parallel imaging to further include the similarities between image supports and phase constraints across coils \cite{Haldar2016}. A further generalization to LORAKS is annihilating filter-based low rank Hankel matrix approach (ALOHA), which extends the finite support constraint to transform domains \cite{Jin2016a}. By relating transform domain sparsity to the existence of annihilating filters in a weighted k-space, where the weighting is determined by the choice of transform domain, ALOHA recasts the reconstruction problem as the low-rank recovery of the associated Hankel matrix.

\section{Machine learning methods for parallel imaging in image space}
\label{sec:imagespace_learning}
The use of machine learning for image-based parallel MR imaging evolves naturally from \Cref{eq:inverse_regularization} based on the following key insights. First, in classic compressed sensing, $\Psi$ are a general regularizers like the image gradient or wavelet transforms, which were not designed specifically with undersampled parallel MRI acquisitions in mind. These regularizers can be generalized to models that have a higher computational complexity. $\Psi$ can be formulated as a convolutional neural network (CNN)~\cite{LeCun1989}, where the model parameters can be learned from training data inspired by the concepts of deep learning~\cite{LeCun2015}. This was already demonstrated earlier in the context of computer vision by Roth and Black~\cite{Roth2009}. They proposed a non-convex regularizer of the following form:
\begin{equation}
    \label{eq:foe_model}
    \Psi{(\mathbf{u})} = \sum_{i=1}^{N_k} \langle \rho_i(\mathbf{K}_i\mathbf{u}),\mathbf{1} \rangle.
\end{equation}
The regularizer in \Cref{eq:foe_model} consists of $N_k$ terms of non-linear potential functions $\rho_i$, $\mathbf{K}_i$ are convolution operators. $\mathbf{1}$ indicates a vector of ones. 
The parameters of the convolution operators and the parametrization of the non-linear potential functions, form the free parameters of the model, which are learned from training data.

\begin{figure}[!t]
	\includegraphics[width = 1 \columnwidth]{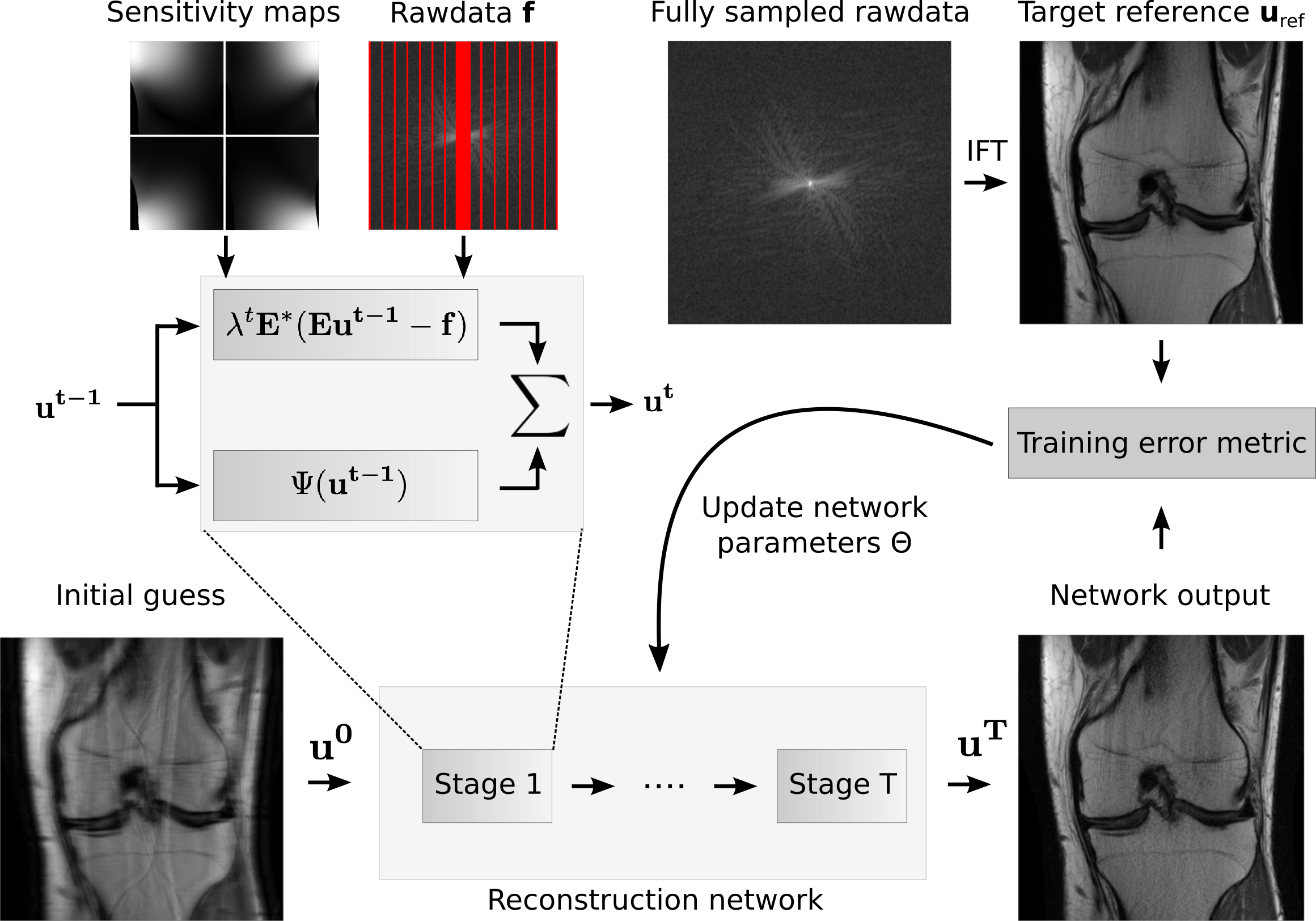}
 	\caption{Illustration of machine learning-based image reconstruction. The network architecture consists of $T$ stages that perform the equivalent of gradient descent steps in a classic iterative algorithm. Each stage consists of a regularizer and a data consistency layer. Training the network parameters $\Theta$ is performed by retrospectively undersampling fully sampled multi-coil raw k-space data and comparing the output of the network $\mathbf{u}^T_s(\Theta)$ to a target reference reconstruction $\mathbf{u}_{\text{ref}}$ obtained from the fully sampled data.}
 	\label{fig:learning_image_space_illustration}
\end{figure}

\begin{figure*}[!t]
\begin{center}
    \includegraphics[width = 1 \textwidth]{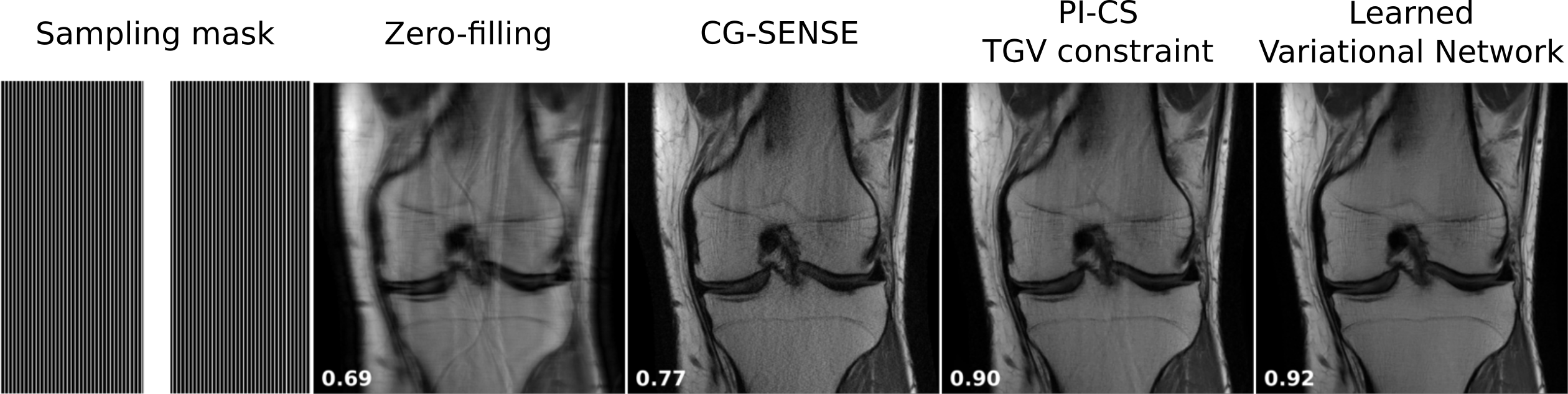}
 	\caption{Comparison of image-domain based parallel imaging reconstructions of a retrospectively accelerated coronal knee acquisition. 
 	The used sampling pattern, zero-filling, CG-SENSE, combined parallel imaging and compressed sensing with a TGV constrained and a learned reconstruction are shown together with their SSIM values to the fully sampled reference. See text in the respective sections for details on the individual experiments.}
 	\label{fig:knee_reconstructions}
 	\vspace{-.3cm}
\end{center}
\end{figure*}

The second insight is that the iterative algorithm that is used to solve \Cref{eq:inverse_regularization} naturally maps to the structure of a neural network, where every layer in the network represents an iteration step of a classic algorithm \cite{Gregor2010}. 
This follows naturally from gradient descent for the least squares problem in \Cref{eq:inverse_leastSquares} that leads to the iterative Landweber method~\cite{Landweber1951}. After choosing an initial $\mathbf{u}^0$, the iteration scheme is given by \Cref{eq:landweber}:
\begin{equation}
    \label{eq:landweber}
	\mathbf{u}^{t} = \mathbf{u}^{t-1} - \alpha^t \mathbf{E}^*(\mathbf{E}\mathbf{u}^{t-1} - \mathbf{f}), \quad t > 0.
\end{equation}
$\mathbf{E}^*$ is the adjoint of the encoding operator and $\alpha^t$ is the step size of iteration $t$. Using this iteration scheme to solve the reconstruction problem in \Cref{eq:inverse_regularization} with the regularizer defined in \Cref{eq:foe_model} leads to the update scheme defined in \Cref{eq:landweber_foe}, which forms the basis of recently proposed image space based machine learning methods for parallel MRI:
\begin{equation}
    \label{eq:landweber_foe}
	\mathbf{u}^{t} = \mathbf{u}^{t-1} - \alpha^t \left( \sum_{i=1}^{N_k} (\mathbf{K}_i)^\top \rho_i'(\mathbf{K}_i \mathbf{u}^{t-1}) + \lambda^t \mathbf{E}^*(\mathbf{E}\mathbf{u}^{t-1} - \mathbf{f}) \right).
\end{equation}
This update scheme can then be represented as a neural network with $T$ stages corresponding to $T$ iteration steps \Cref{eq:landweber_foe}. $\rho_i'$ are the first derivatives of the nonlinear potential functions $\rho_i$, which are represented as activation functions in the neural network. The transposed convolution operations ${\mathbf{K}}_i^\top$ correspond to convolutions with filter kernels rotated by 180 degrees. Most recently proposed approaches follow this structure, and their difference mainly lies is the used model architecture. The idea of the variational network~\cite{Hammernik2018} follows the structure of classic variational methods and gradient-based optimization, and the network architecture is designed to mimic a classic iterative image reconstruction. The approach from Aggarwal et al.~\cite{Aggarwal2019} follows a similar design concept, while using convolutional neural networks (CNNs), but shares the same set of parameters for all stages of the network, thus reducing the number of free parameters. It also uses an unrolled conjugate-gradient step for data consistency instead of the gradient based on in \Cref{eq:landweber}. 

An illustration of image space based machine learning for parallel MRI along the lines of~\cite{Hammernik2018,Aggarwal2019} is shown in \Cref{fig:learning_image_space_illustration}. 
To determine the model parameters of the network that will perform the parallel imaging reconstruction task, an optimization problem needs to be defined that minimizes a training objective. In general, this can be formulated in a supervised or unsupervised manner. Supervised approaches are predominantly used while unsupervised approaches are still a topic of ongoing research (an approach for low-dose CT reconstruction was presented in~\cite{Wu2017}). Therefore, we will focus on supervised approaches for the remainder of this section. We define the number of stages, corresponding to gradient steps in the network, as $T$. $s$ is the current training image out of the complete set of training data $S$. The variable $\Theta$ contains all trainable parameters of the reconstruction model. The training objective then takes the following form:
\begin{equation}
  \label{eq:image_space_learning_training}
  L(\Theta) = \min_{\Theta} \frac{1}{2S} \sum_{s=1}^{S} \| \mathbf{u}^T_s(\Theta) - \mathbf{u}_{\text{ref},s} \|_2^2.
\end{equation}

As it is common in deep learning, \Cref{eq:image_space_learning_training} is a non-convex optimization problem that is solved with standard numerical optimizers like stochastic gradient descent or ADAM~\cite{Kingma2014}. This requires the computation of the gradient of the training objective with respect to the model parameters $\Theta$. This gradient can be computed via backpropagation~\cite{LeCun2012}:
\begin{equation}
	\frac{\partial L(\Theta)}{\partial \Theta^t} = \frac{\partial \mathbf{u}^{t+1}}{\partial \Theta^t} \cdot \frac{\partial \mathbf{u}^{t+2}}{\partial \mathbf{u}^{t+1}} \hdots \cdot \frac{\partial \mathbf{u}^T}{\partial \mathbf{u}^{T-1}} \cdot \frac{\partial L(\Theta)}{\partial \mathbf{u}^T}.
\end{equation}


The basis of supervised approaches is the availability of a target reference reconstruction $\mathbf{u}_{\text{ref}}$. This requires the availability of a fully-sampled set of raw phased array coil k-space data. This data is then retrospectively undersampled by removing k-space data points as defined by the sampling trajectory in the forward operator $\mathbf{E}$ and serves as the input of the reconstruction network. The current output of the network $\mathbf{u}^T_s(\Theta)$ is then compared to the reference $\mathbf{u}_{\text{ref}}$ via an error metric. The choice of this error metric has an influence on the properties of the trained network, which is a topic of currently ongoing work. A popular choice is the mean squared error (MSE), which was also used in \Cref{eq:image_space_learning_training}. Other choices are the $\ell_1$ norm of the difference~\cite{Hammernik2017b} and the structural similarity index (SSIM)~\cite{Wang2004}. Research on generative adversarial networks~\cite{Goodfellow2014} and learned content loss functions is currently in progress. The current literature in this area is further noted in the discussion section.

An example reconstruction that compares the variational network learning approach from~\cite{Hammernik2018} to CG-SENSE and constrained reconstructions from the previous sections is shown in \Cref{fig:knee_reconstructions} together with the SSIM to the fully sampled reference. It can be observed that the learned reconstruction outperforms the other approaches in terms of artifact removal and preservation of small image features, which is also reflected in the highest SSIM. All source code\footnote{\url{https://github.com/VLOGroup/mri-variationalnetwork}} for this method is available online.

\section{Machine learning methods for parallel imaging in k-space} \label{sec:kspace_learning}

There has been a recent interest in using neural network to improve the k-space interpolation techniques using non-linear approaches in a data-driven manner. These newer approaches can be divided into two groups based on how the interpolation functions are trained. The first group uses scan-specific ACS lines to train neural networks for interpolation, similar to existing interpolation approaches, such as GRAPPA or non-linear GRAPPA. The second group uses training databases, similar to the machine learning methods discussed in image domain parallel imaging. 

\begin{figure}[!t]
\centering
	\includegraphics[width =\columnwidth]{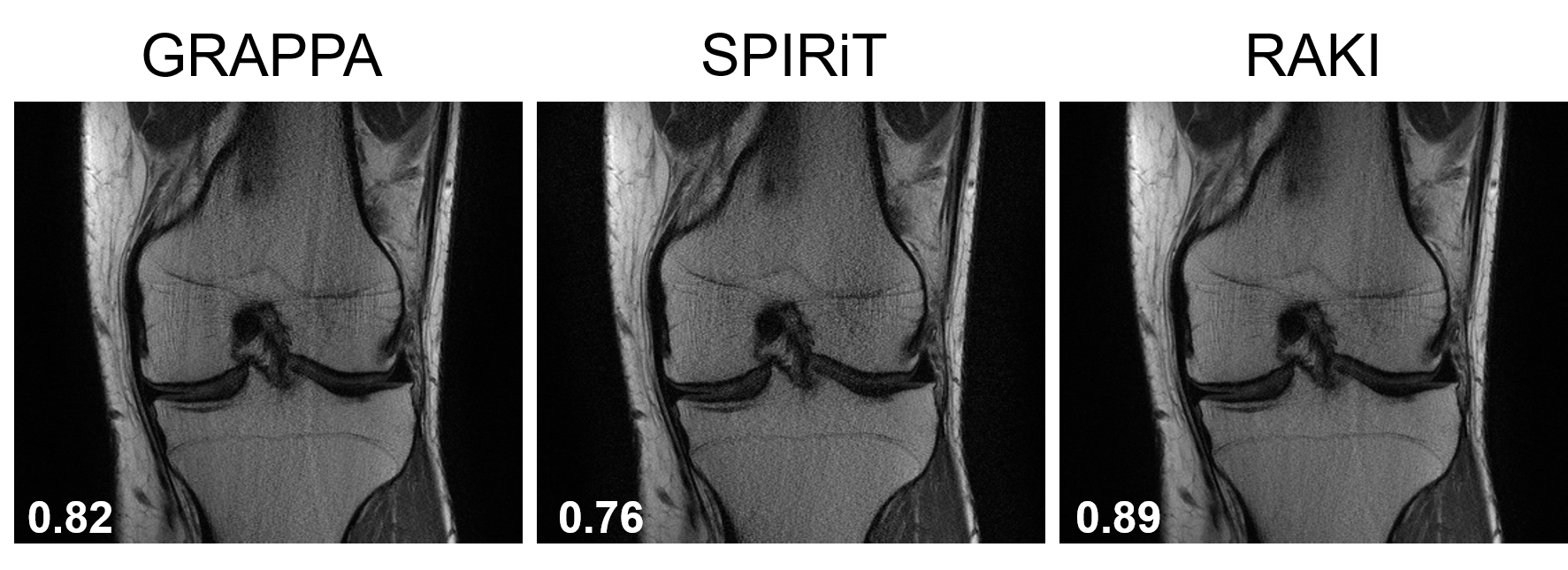}
 	\caption{Comparison of k-space parallel imaging reconstructions of a retrospectively accelerated coronal knee acquisition, as in \Cref{fig:knee_reconstructions}. Due to the small size of the ACS data relative to the acceleration rate, the methods, none of which utilizes training databases, exhibit artifacts. GRAPPA has residual aliasing, whereas SPIRiT shows noise amplification. These are reduced in RAKI, though the residual artifacts remain. Respective SSIM values reflect these visual assessment.}
 	\label{fig:knee_kspace}
 	\vspace{-.2cm}
\end{figure}

Robust artificial-neural-networks for k-space interpolation (RAKI) is a scan-specific machine learning approach for improved k-space interpolation \cite{Akcakaya2019}. This approach trains CNNs on ACS data, and uses these for interpolating missing k-space points from acquired ones. The interpolation function can be represented by
\begin{align}
     {\kappa}&_j(k_x, k_y - m\Delta k_y) = f_{j,m}(\{{\kappa}_c(k_x - b_x \Delta_x, 
     \nonumber \\
     &\quad k_y - R b_y \Delta_y)\}
     _{b_x \in [-B_x,B_x], b_y \in [-B_y, B_y], c \in [1, n_c]}),
\end{align}
where $f_{j,m}$ is the interpolation rule implemented via a multi-layer CNN for outputting the k-space of the $m^\textrm{th}$ set of uniformly spaced missing lines in the $j^\textrm{th}$ coil, $R$ is the undersampling rate, $B_x, B_y$ are parameters specified by the receptive field of the CNN, $n_c$ is the number of coils. 
Thus, the premise of RAKI is similar to GRAPPA, while the interpolation function is implemented using CNNs, whose parameters are learned from ACS data with an MSE loss function. The scan-specific nature of this method is attractive since it requires no training databases, and can be applied to scenarios where a fully-sampled gold reference cannot be acquired, for instance in perfusion or real-time cardiac MRI, or high-resolution brain imaging. 
\begin{figure}
\centering
	\includegraphics[width = \columnwidth]{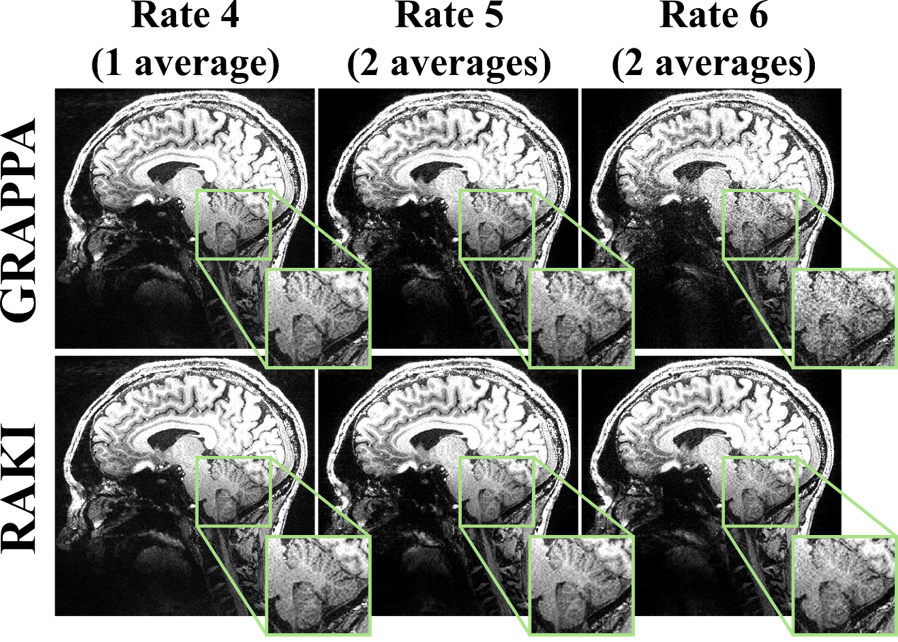}
	 \caption{A slice from a high-resolution (0.6 mm isotropic) 7T brain acquisition, where all acquisitions were performed with prospective acceleration. It is difficult to acquire fully-sampled reference datasets for training for such acquisitions, thus two scan-specific k-space methods were compared. The CNN-based RAKI method visibly reduced noise amplification compared to the linear GRAPPA reconstruction.}
 	\label{fig:raki7T}
 	\vspace{-.2cm}
\end{figure}
Example RAKI and GRAPPA reconstructions for such high-resolution brain imaging datasets, which were acquired with prospective undersampling are shown in \Cref{fig:raki7T}. These data were acquired on a 7T system (Siemens Magnex Scientific) with 0.6 mm isotropic resolution. $R = 5,6$ data were acquired with two averages for improved SNR to facilitate visualization of any residual artifacts. Other imaging parameters are available in \cite{Akcakaya2019}. For these datasets, RAKI leads to a reduction in noise amplification compared to GRAPPA. Note the noise reduction here is based on exploiting properties of the coil geometry, and not on assumptions about image structure, as in traditional regularized inverse problem approaches, as in \Cref{sec:2c}. 
However, the scan-specificity also comes with downsides, such as the computational burden of training for each scan \cite{Zhang2018a}, as well as the requirement for typically more calibration data. In \Cref{fig:knee_kspace}, reconstructions of the knee dataset from \Cref{fig:knee_reconstructions} are shown, where all methods, which rely on subject-specific calibration data, exhibit a degree of artifacts, due to the small size of the ACS region, while RAKI has the highest SSIM among these.

\begin{figure*}[!t]
\centering
	\includegraphics[width = \textwidth]{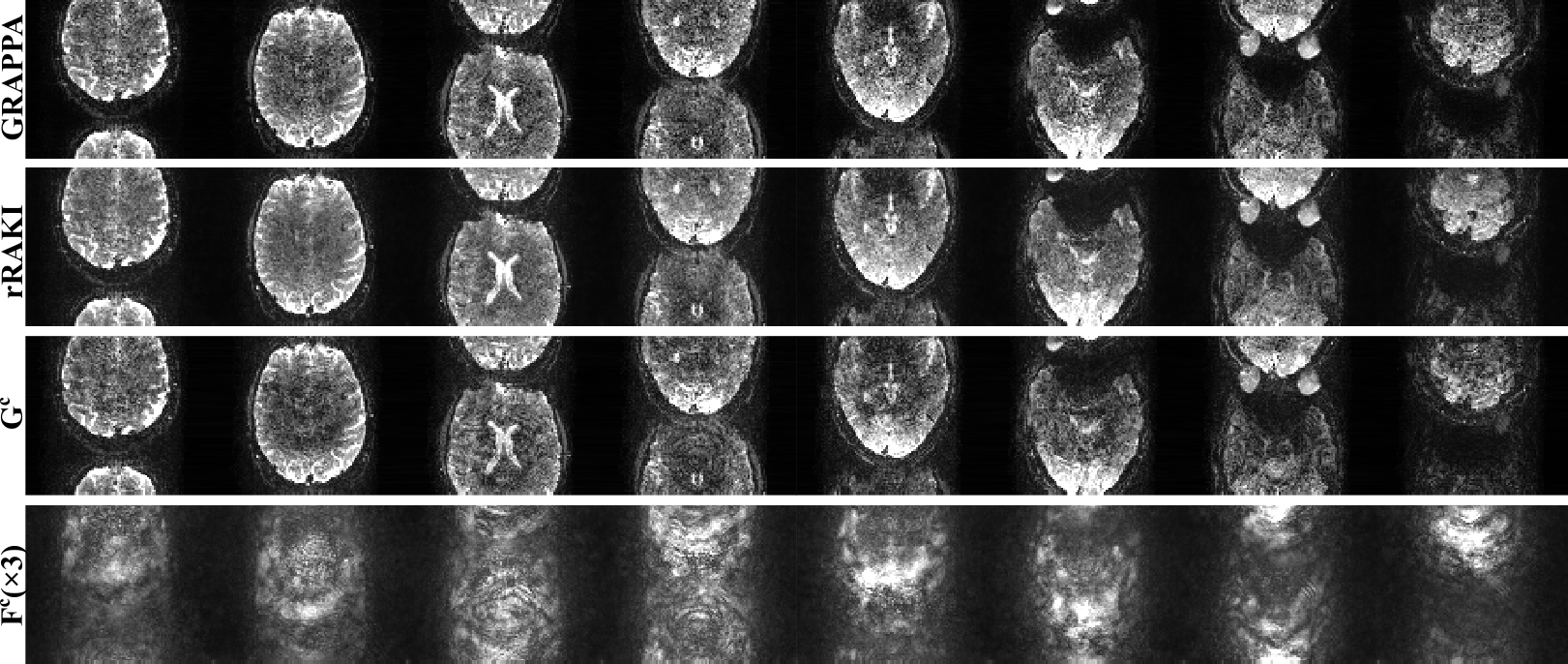}
 	\caption{Reconstruction results of simultaneous multi-slice imaging of 16 slices in fMRI, where the central 8 slices are shown. GRAPPA method exhibits noise amplification at this high acceleration rate. The rRAKI method, whose linear and residual components are depicted by $G^C$ and $F^C$ respectively, exhibits reduced noise. Due to imperfections in the ACS data for this application, the residual component includes both noise amplification and residual artifacts.}
 	\label{fig:rraki}
\end{figure*}

While originally designed for uniform undersampling patterns, this method has been extended to arbitrary sampling, building on the self-consistency approach of SPIRiT \cite{Hosseini2019}. Additionally, recent work has also reformulated this interpolation procedure as a residual CNN, with residual defined based on a GRAPPA interpolation kernel \cite{Zhang2019}. Thus, in this approach called residual RAKI (rRAKI), the CNN effectively learns to remove the noise amplification and artifacts associated with GRAPPA, giving a physical interpretation to the CNN output, which is similar to the use of residual networks in image denoising \cite{Zhang2017denoise}. An example application of the rRAKI approach in simultaneous multi-slice imaging \cite{Zhang2018b} is shown in \Cref{fig:rraki}.

\begin{figure}[!b]
\centering
	\includegraphics[width =\columnwidth]{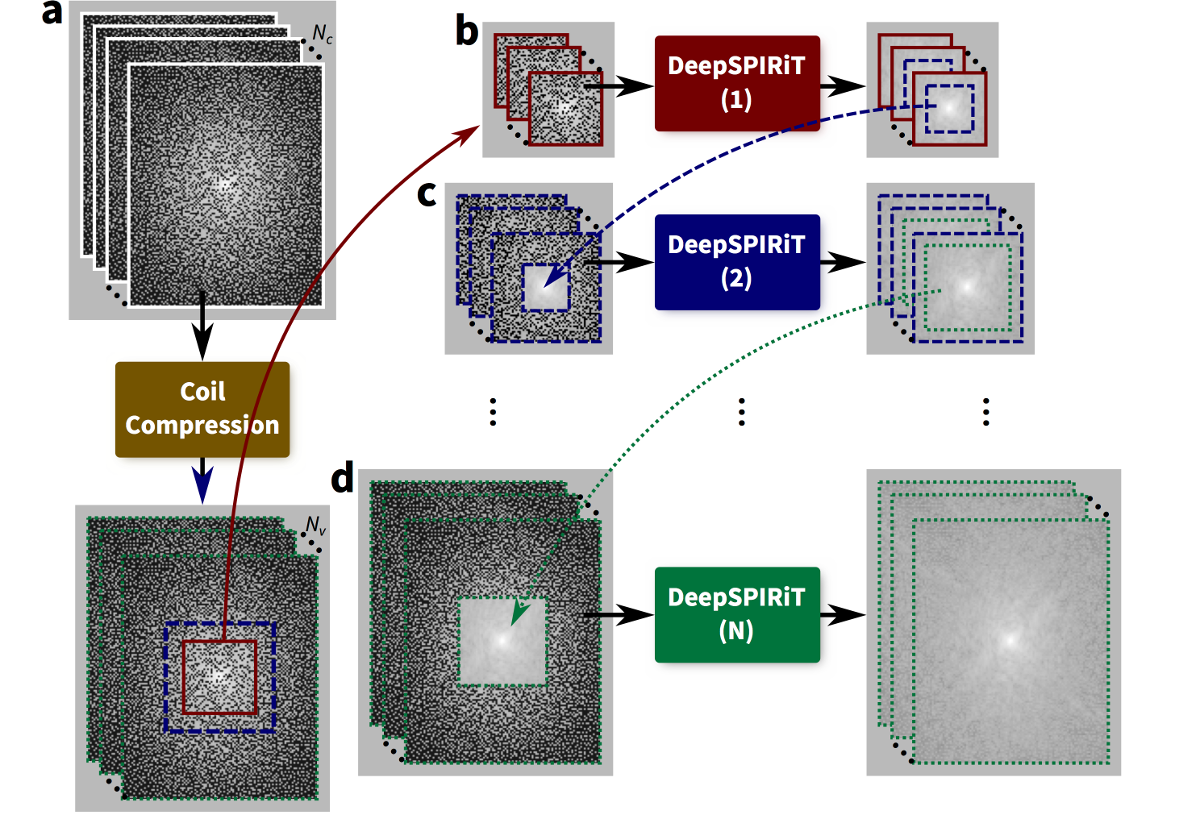}
	\caption{The multi-resolution k-space interpolation in DeepSPIRiT uses distinct CNNs for diffrent regions of k-space, successively refining the resolution of the reconstructed k-space.}
 	\label{fig:deepspirit}
\end{figure}

A different line of work, called DeepSPIRiT, explores using CNNs trained on large databases for k-space interpolation with a SPIRiT-type approach \cite{Cheng2018}. Since sensitivity profiles and number of coils vary for different anatomies and hardware configurations, k-space data in the database were normalized using coil compression to yield the same number of channels \cite{Buehrer2007, Huang2008}. Coil compression methods effectively capture most of the energy across coils in a few virtual channels, with the first virtual channel containing most of the energy, the second being the second most dominant, and so on, in a manner reminiscent of principal component analysis. After this normalization of the k-space database, CNNs are trained for different regions of k-space, which are subsequently applied in a multi-resolution approach, successively improving the resolution of the reconstructions, as illustrated in \Cref{fig:deepspirit}. The method was shown to remove aliasing artifacts, though difficulty with high-resolution content was noted. Since DeepSPIRiT trains interpolation kernels on a database, it does not require calibration data for a given scan, potentially reducing acquisition time further. 

Neural networks have also been applied to the Hankel matrix based approaches in k-space \cite{Han2018a}. Specifically, the completion of the weighted k-space in ALOHA method has been replaced with a CNN, trained with an MSE loss function. The method was shown to not only improve the computational time, but also the reconstruction quality compared to original ALOHA by exploiting structures beyond low-rankness of Hankel matrices.

\section{Discussion}
\subsection{Issues and open problems}
Several advantages of machine learning approaches over classic constrained reconstruction using predefined regularizers have been proposed in the literature. First, the regularizer is tailored to a specific image reconstruction task, which improves the removal of residual artifacts. This becomes particularly relevant in situations where the used sampling trajectory does not fulfill the incoherence requirements of compressed sensing, which is often the case for clinical parallel imaging protocols. Second, machine learning approaches decouple the compute-heavy training step from a lean inference step. In medical image reconstruction, it is critical to have images available immediately after the scan, while prolonged training procedures that can be done on specialized computing hardware, are generally acceptable. The training for the experiment in \Cref{fig:knee_reconstructions} took 40 hours for 150 epochs with 200 slices of training data on a single NVIDIA M40 GPU with 12GB of memory. Training data, model and training parameters exactly follow the training from~\cite{Knoll2019}.  Reconstruction of one slice then took 200ms, in comparison to 10ms for zero filling, 150ms for CG-SENSE and 10000ms for the PI-CS TGV constrained reconstruction.

The focus in \Cref{sec:imagespace_learning} and \Cref{sec:kspace_learning} were on methods that were developed specifically in the context of parallel imaging. Some architectures for image domain machine learning have been designed specifically towards a target application, for example dynamic imaging~\cite{Schlemper2018,Qin2019}. In their current form, these were not yet demonstrated in the context of multi-coil data. The approach recently proposed by Zhu et al. learns the complete mapping from k-space raw data to the reconstructed image~\cite{Zhu2018a}. The proposed advantage is that since no information about the acquisition is included in the forward operator $A$, it is more robust against systematic calibration errors during the acquisition. This comes at the price of a significantly higher number of model parameters. The corresponding memory requirements make it challenging use this model for matrix sizes that are currently used in clinical applications. 
We also note that there are fewer works in k-space machine learning methods for MRI reconstruction. This may be due to the different nature of k-space signal that usually has very different intensity characteristics in the center versus the outer k-space, which makes it difficult to generalize the plethora of techniques developed in computer vision and image processing that exploit properties of natural images.

Machine learning reconstruction approaches also come with a number of drawbacks when compared to classic constrained parallel imaging. First, they require the availability of a curated training data set that is representative so that the trained model generalized towards new unseen test data. While recent approaches from the literature~\cite{Hammernik2018,Aggarwal2019,Schlemper2018,Qin2019,Chen2018} have either been trained with hundreds of examples rather than millions of examples as it is common in deep learning for computer vision~\cite{Deng2009,LeCun2015}, or trained on synthetic non-medical data~\cite{Zhu2018a} that is publicly available from existing databases. However, this is still a challenge that will potentially limit the use of machine learning to certain applications. Several applications in imaging of moving organs, such as the heart, or in imaging of the brain connectivity, such as diffusion MRI, cannot be acquired with fully-sampled data due to constraints on spatio-temporal resolutions. This hinders the use of fully-sampled training labels for such datasets, highlighting applications for scan-specific approaches or unsupervised training strategies.

These reconstruction methods also require the availability of computing resources during the training stage. This is a less critical issue due to the increased availability and reduced prices of GPUs. The experiments in this paper were made with computing resources that are available for less than 10,000 USD, which are usually available in academic institutions. In addition, the availability of on-demand cloud based machine learning solutions is constantly increasing. 

A more severe issue is that in contrast to conventional parallel imaging and compressed sensing, machine learning models are mostly non-convex. Their properties, especially regarding their failure modes and generalization potential for daily clinical use, are understood less well than conventional iterative approaches based on convex optimization. For example, it was recently shown that while reconstructions generalize well with respect to changes in image contrast between training and test data, they are susceptible towards systematic deviations in SNR~\cite{Knoll2019}. It is also still an open question how specific trained models have to be. 
Is it sufficient to train a single model for all types of MR exams, or are separate models required for scans of different anatomical areas, pulse sequences, acquisition trajectories and acceleration factors as well as scanner manufacturers, field strengths and receive coils? \cite{icassp2017_open_problems} 
While pre-training a large number of separate models for different exams would be feasible in clinical practice, if certain models do not generalize with respect to scan parameter settings that are usually tailored to the specific anatomy of an individual patient by the MR technologist, this will severely impact their translational potential and ultimately their clinical use.

Finally, the choice of the loss function that is used during the training has an impact on the properties of the trained network, and a particular ongoing research direction is the use of GANs~\cite{Goodfellow2014,Arjovsky2017,Gulrajani2017,Mao2017}. This is an interesting direction because these models have the potential to create images that are visually indistinguishable from fully sampled reference images, where features in the images that are not supported by the amount of acquired data, are hallucinated by the network. This situation obviously must be avoided in any application in medical imaging. Strategies to mitigate this effect are the combination of GANs with conventional error metrics like MSE~\cite{Isola2016image,Ledig2017c}. Comparable approaches were used in the context of MRI reconstruction~\cite{Quan2017,Shitrit2017,Yang2018,Hammernik2018a,Kim2018a,Mardani2018}. 

\subsection{Availability of training databases and community challenges}
As mentioned in the previous section, one open issue in the field of machine learning reconstruction for parallel imaging is the lack of publicly available databases of multi-channel raw k-space data. This restricts the number of researchers who can work in this field to those who are based at large academic medical centers where this data is available, and for the most part excludes the core machine learning community that has the necessary theoretical and algorithmic background to advance the field. In addition, since the used training data becomes an essential part of the performance of a certain model, it is currently almost impossible to compare new approaches that are proposed in the literature with each other if the training data is not shared when publishing the manuscript. While the momentum in initiatives for public releases of raw k-space data is growing~\cite{fastMRIcommunity}, the number of available data sets is still on the order of hundreds and limited to very specific types of exams. Examples of publicly available rawdata sets are mridata.org\footnote{\url{https://mridata.org}} and the fastMRI dataset\footnote{\url{https://fastmri.med.nyu.edu/}}.

\section{Conclusion}

Machine learning methods have recently been proposed to improve the reconstruction quality in parallel imaging MRI. These techniques include both image domain approaches for better image regularization and k-space approaches for better k-space completion. While the field is still in its development, there are many open problems and high-impact applications, which are likely to be of interest to the broader signal processing community.

\bibliographystyle{IEEEbib}



\end{document}